\documentclass[a4paper,10pt]{article}
\usepackage{geometry,color,graphicx}
\usepackage{amssymb,amsmath}
\usepackage{graphicx}
\numberwithin{equation}{section}

\begin{document}
\title{ Thermodynamic Geometry of Reissener-Nordstr\"{o}m-de Sitter black hole and its extremal case}
\author{{R. Tharanath \thanks{Email: tharanath.r@gmail.com} ,
\hspace{1mm} Jishnu Suresh, \hspace{1mm}
Nijo Varghese, \hspace{1mm}
and  V C Kuriakose} \\\\
\small{\em{Department of Physics, Cochin University of Science and Technology, Cochin 682022, Kerala, India}} }

\date{}
\maketitle

\begin{abstract}
We study the thermodynamics and the different  thermodynamic geometric methods   
of Reissener-Nordstr\"{o}m-de Sitter black hole and its extremal  
case, which is similar to the de Sitter black hole coupled to a scalar field,
rather called an MTZ black hole. While studying the thermodynamics of the systems, we could find  some abnormalities. 
In both cases, the thermodynamic geometric methods could  give the correct explanation for the all abnormal thermodynamic behaviors in the system. 

\end{abstract}

\section{\label{sec:intro}Introduction}

Black hole thermodynamics(BHT) is receiving more and more attentions in recent times. 
The path breaking findings of 
 Hawking and Bekenstein \cite{hawking,bekenstein,bekenstein2}
 made in 1970s helped us to think that black holes are not truly black but are thermal objects emitting radiations.
 The main reason for the interest shown in BHT is that it unites quantum theory, gravitation and thermodynamics 
 and it may open a way to quantum formulation of gravity. 
The normal BHT is evolved by considering the black hole mass as one of the thermodynamic potentials
and assuming the area law straightforwardly. We have usual thermodynamic expressions to derive the quantities like temperature,
entropy, heat capacity etc.  In the thermodynamic studies,  heat capacity is quite important because it can tell upon the thermodynamic stability of systems.
 For example, we could easily find the heat capacity of  Schwarzschild black hole where 
one can see that it is negative and hence we may conclude that Schwarzschild black hole is thermodynamically unstable. In a similar fashion we can extend this 
stability analysis to other black holes 
and we could see that certain black holes show both positive and negative  heat capacities. The transition is through an infinite discontinuity in the heat capacity. 
Such a phase transition is called a second order thermodynamic phase transition in BHT.

While closely looking at the normal thermodynamics of black holes, in many cases
we could not identify the exact reasons for the abnormalities of energy(mass), temperature and heat capacity shown by the system.
During the last few decades many attempts have been made to introduce different geometric concepts into ordinary thermodynamics. 
Hermann \cite{hermann} formulated the concept of thermodynamic phase space as a differential manifold with a natural contact structure. 
In the thermodynamic phase space there exists a special subspace of thermodynamic equilibrium states. 

In 1975 Weinhold\cite{weinhold} introduced an alternate geometric method in which a metric is introduced in the space of equilibrium states of thermodynamic systems. 
There he used the idea of conformal mapping from the Riemannian space to thermodynamic space, in which the new metric could be identified in terms of the
thermodynamic potentials. Weinhold metric is given by
\begin{equation}
 g^{W}_{ij}=\partial_i \partial_j U(S,N^r),
\end{equation}

where $S$ is the entropy, $U$ is the internal energy and $N^r$ denotes other extensive variables of the system . There are systems which  Weinhold metric 
gives correct explanation.
In an attempt to formulate the concept of thermodynamic length, in 1979 Ruppeiner \cite{ruppeiner} 
 introduced another metric which is conformaly equivalent to Weinhold's metric. 
The Ruppeiner metric( which is the minus signed Hessian in entropy representation) is given by 
\begin{equation}
 g^{R}_{ij}=-\partial_i \partial_j S(M,N^r).
\end{equation}
The Ruppiner geometry is conformaly related to the Weinhold geometry by \cite{mrugala,salamon}
\begin{equation}
 ds_{R}^{2}=\frac{1}{T} ds^{2}_{W}
\end{equation}
where $T$ is the temperature of the system under consideration. Since the proposal of Weinhold, a number of investigations 
have been done to analyze the thermodynamic
geometry of various thermodynamic systems. The Weinhold and Ruppeiner geometries have also been analyzed for several
black holes to find out the thermodynamical abnormalities\cite{ferr,jjkb,jjka,am,ama,aman,scws,cc,sst,med,mz}.

Geometrothermodynamics(GTD)\cite{q1,q2,q3} is the latest attempt in this direction. In order to describe a thermodynamic system with $n$ degrees of freedom, 
we consider in GTD, a thermodynamic phase space which
is defined mathematically as a Riemannian contact
manifold$(\mathcal{T} , \Theta, G)$, where $\mathcal{T}$ is a $2n + 1$ dimensional manifold, 
$\Theta$ defines a contact structure on $\mathcal{T}$ and $G$ is a Legendre invariant metric on $\mathcal{T}$.
The pair $( \mathcal{T}, \Theta)$ is called a contact manifold\cite{hermann} only if $\mathcal{T}$ is
differentiable and $\Theta$ satisfies the condition $\Theta \wedge (d \Theta)^{n} \neq 0$;
which actually preserves the essential Legendre invariance while making the conformal transformations.
The space of equilibrium states is an $n$ dimensional manifold $(\mathcal{E}, g)$, where $\mathcal{E} \subset \mathcal{T}$ is defined by a 
smooth mapping
$\phi : \mathcal{E} \rightarrow \mathcal{T}$ such that the pullback $\phi^* (\Theta) = 0$, and a
Riemannian structure $g$ is induced naturally in the $\mathcal{E}$ by
means of $g = \phi^{(G)} $. It is then expected in GTD that
the physical properties of a thermodynamic system in
a state of equilibrium can be described in terms of the
geometric properties of the corresponding space of equilibrium states $\mathcal{E}$. The smooth mapping can be read in
terms of coordinates as, $\phi:(E^{a}) \rightarrow (\Phi, E^{a} , I^{a} )$ with
$\Phi$ representing the thermodynamic potential, $E^{a}$ and
$I^{a}$ representing the extensive and intensive thermodynamic
variables respectively if the condition $\phi^* (\Theta) = 0$ is satisfied, i.e.,

\begin{equation}
  d\Phi = \delta_{ab}I^{a}dE^{b}\ , \quad \frac{\partial \Phi}{\partial E^{a}}=\delta_{ab}I^{b}\ .
\label{firstlaw}
\end{equation}
The first of these equations corresponds to the first law of thermodynamics, whereas the second one is usually 
known as the condition for thermodynamic equilibrium\cite{callen}.

Legendre invariance guarantees that the geometric
properties of $G$ do not depend on the thermodynamic
potential used in its construction. Hernavo Quevedo\cite{q1} introduced the idea and constructed a general form for
the Legendre invariant metric. The general choice of GTD metric is as follows

\begin{equation}
 g=\phi^*(G)= \left(E^{c}\frac{\partial{\Phi}}{\partial{E^{c}}}\right)
\left(\eta_{ab}\delta^{bc}\frac{\partial^{2}\Phi}{\partial {E^{c}}\partial{E^{d}}} dE^a dE^d \right).
\end{equation}

The thermodynamic geometry of a black hole is still a most fascinating subject and there are many unresolved issues in BHT. Using this we could 
solve a number of issues related to the abnormal behaviour of mass, temperature and heat capacity. The metric is built up from a Legendre
invariant thermodynamic potential and from its first and second order partial derivatives with respect to the extensive
variables. The earlier studies show that the thermodynamic stability of systems depend on the potential we have chosen.
This contradiction has been removed by using new Legendre invariant metric introduced in the GTD. In this
paper we describe the phase transition in terms of curvature singularities.

The main purpose of the present work is to show that
the thermodynamic geometric methods can be used to explain the thermodynamics of RNdS black hole and MTZ black hole,
in both cases we are taking the
cosmological constant as an extensive variable. The organization of the manuscript is as follows, in section[2] we
study the usual thermodynamics of RNdS black hole.
We analyze the thermodynamic geometry of RNdS as
well. The extremal case of RNdS or the MTZ black hole
has been studied in section [3] followed by its thermodynamic geometry. Section [4] is devoted to conclusion and
discussions.

\section{ RN de Sitter Black hole}
\subsection{Thermodynamics}

The Reissner-Nordstr\"{o}m-de Sitter solution describes a static, spherically symmetric black hole carrying mass $M$, charge $Q$ and a non vanishing
cosmological constant $\Lambda$.

\begin{figure}[h]
\includegraphics[scale=0.6]{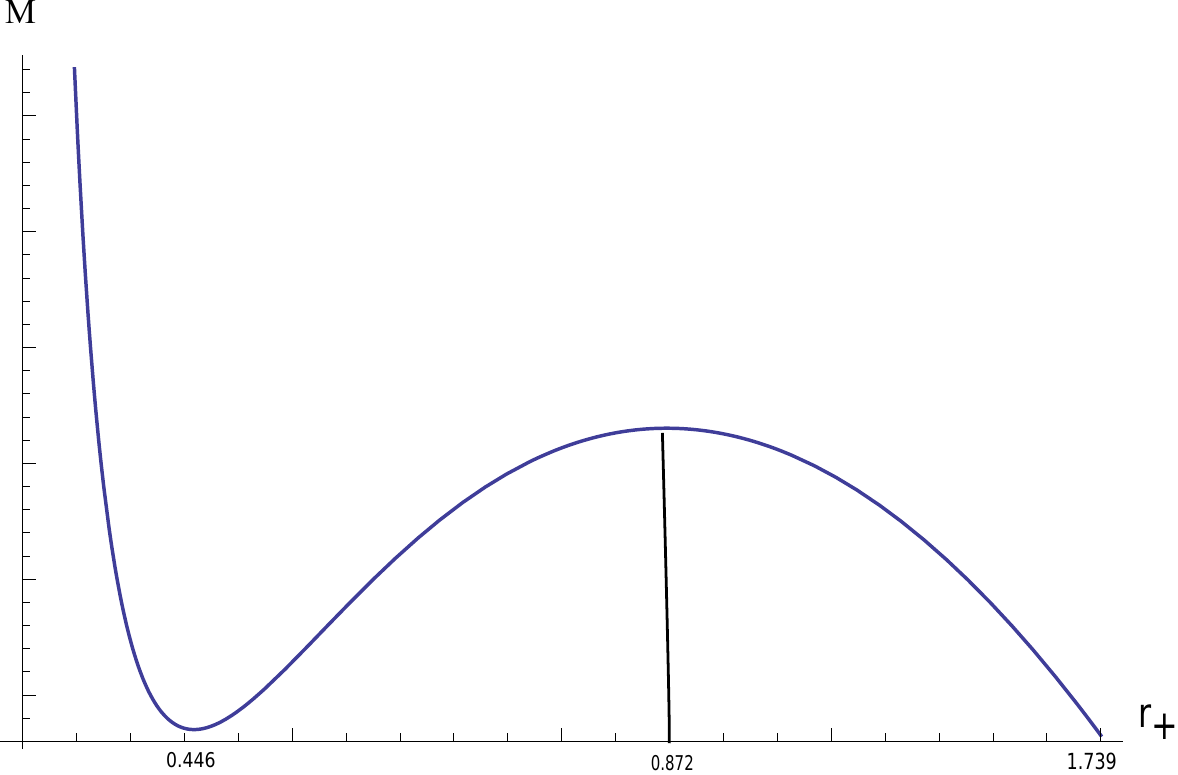}
\caption{\label{fig:orlfo} Variation of mass with respect to horizon radius $r_+$. We set $\alpha=\sqrt{3}$ and $Q=0.4$.}
\end{figure}

The metric of of the RNdS black hole is written as
\begin{equation}
 ds^{2}=f(r)dt^{2}-f(r)^{-1}dr^{2}-r^{2}(d\Omega^{2}),
\end{equation}
where $f(r)$ is equal to

\begin{figure}[h]
\includegraphics[scale=0.28]{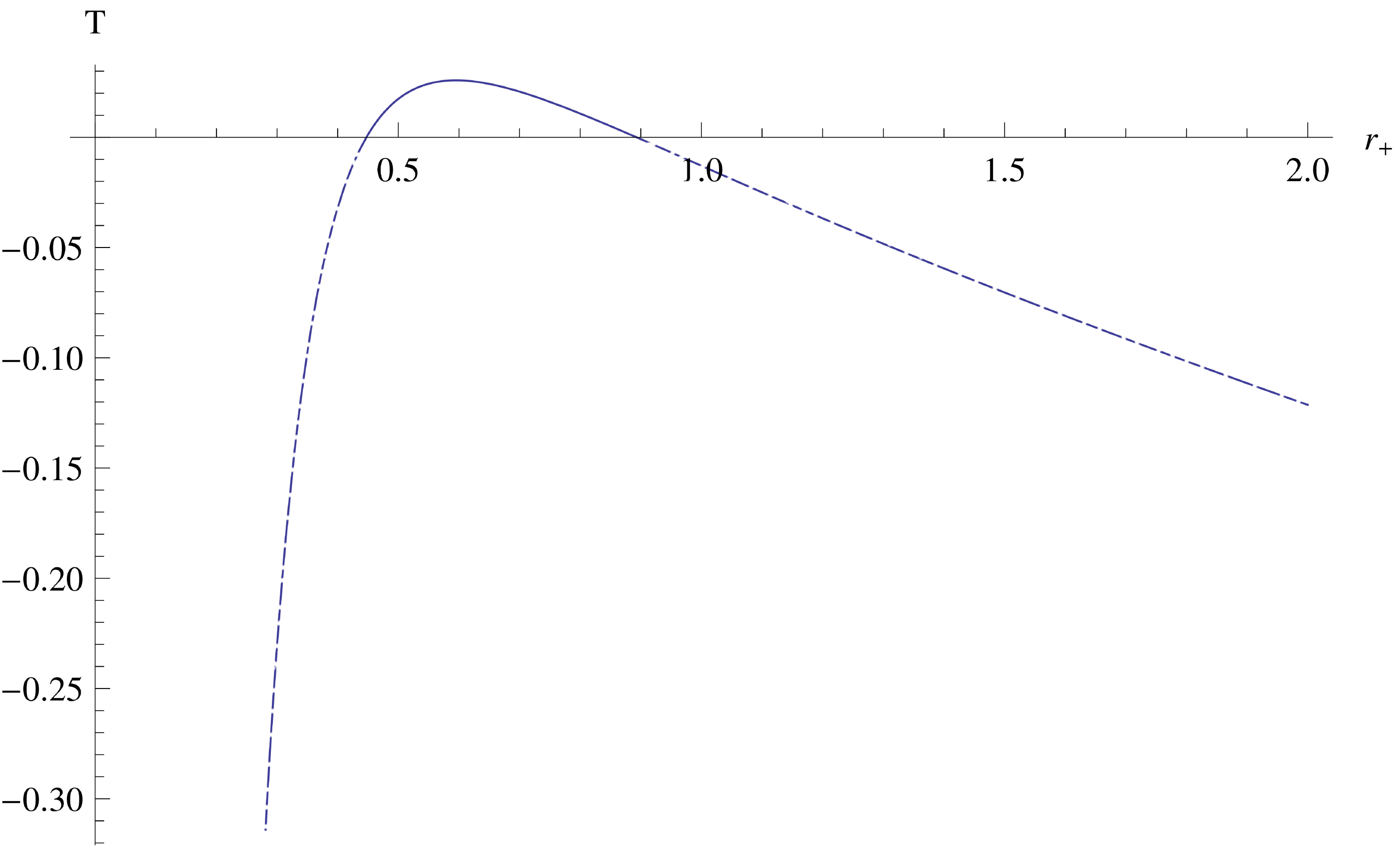}
\caption{\label{fig:orlfo} Variation of temperature with respect to horizon radius $r_+$. We set $\alpha=\sqrt{3}$ and $Q=0.4$. The dashed portion is the negative - unphysical regime of temerature.}
\end{figure}

 \begin{equation}
  f(r)= 1-\frac{2M}{r}+ \frac{Q^2}{r^2}-\frac{\Lambda r^2}{3},
 \end{equation}
here we could find the mass of the black hole  $M$, in terms of its entropy $S$, charge $Q$ and the radius of curvature 
of the de Sitter space $\alpha$, where $\alpha$ is connected to the cosmological constant  $\Lambda$ through the relation
\begin{equation}
 \Lambda= \frac{(N-1)(N-2)}{2 \alpha^2},
\end{equation}

   where $N$ is the dimension of the space time. 
  Using the relation between entropy $S$ and event horizon radius $r_{+}$, $S= \pi r_+^2$ (from area law), 
  we can write the mass term as,
\begin{equation}
 M(S,Q,\alpha)=\frac{\pi^2 Q^2 \alpha^2 + \pi S \alpha^2 - S^2}{2 \pi^{\frac{3}{2}} \alpha^2 S^{\frac{1}{2}}}.
\end{equation}

The other thermodynamic parameters can be identified using the above expression of mass as
$T=\frac{\partial M}{\partial S}, ~~~ C= T \frac{\partial S}{\partial T}$
and hence we can obtain the temperature as a function of $S$ and $Q$,
\begin{equation}
 T= \frac{ \alpha^2 \pi (S - \pi Q^2)- 3 S^2}{4 \alpha^2 \pi^{\frac{3}{2}}S^{\frac{3}{2}}}.
\end{equation}

\begin{figure}[h]
\includegraphics[scale=0.8]{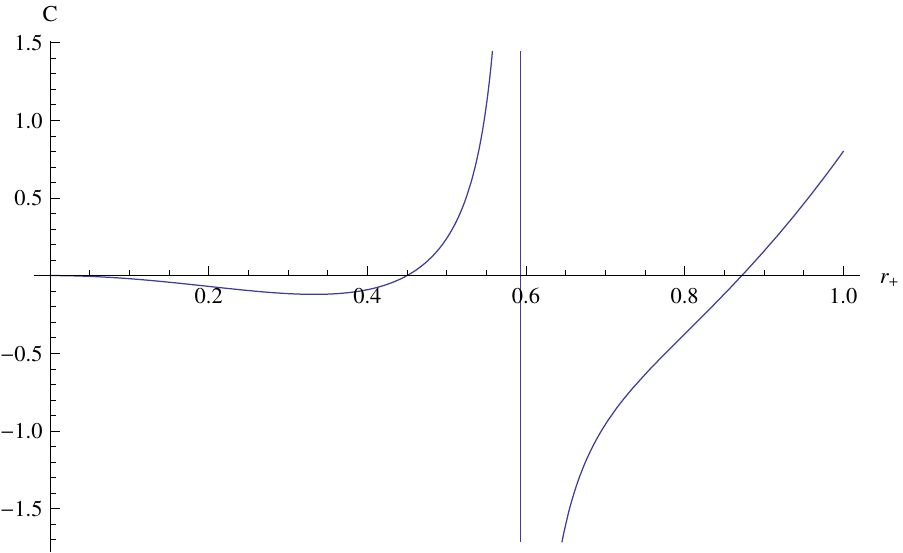}
\caption{\label{fig:orlfo} Variation of heat capacity with respect to horizon radius $r_+$. We set $\alpha=\sqrt{3}$ and $Q=0.4$.}
\end{figure}

The heat capacity is also obtained as
\begin{equation}
 C=\frac{2S[\alpha^2 \pi ( \pi Q^2 -S) + 3 S^2 ]}{3 S^2 + \alpha^2 \pi (S- 3 \pi Q^2)}.
\end{equation}

Here we have obtained three thermodynamic quantities: mass$(M)$, temperature$(T)$ and specific heat$(C)$ and we have plotted all of them in tems of horizon radius
$r_+$( Figs 1-3). We could easily identify the abnormal behaviors in each of these quantities
from their plots. In Fig.1, we could see that mass of the black
hole become zero at two places and reaches a maximum value at a particular value of $r_+$. 
Temperature is positive only in between a certain range of horizon radius $r_+$, which is shown in Fig.2; the negative temperature is
one of the issues we need to find a satisfactory explanation. 
Finally, the heat capacity changes from negative(unstable)
phase to positive(stable) phase through an infinite discontinuity, as in Fig.3. It refers to a second order thermodynamic
phase transition in BHT. Thus the the black hole will be thermodynamically stable for only a certain range of values of mass, temperature and heat capacity.

On a close examination of Fig.1
 we can see that mass becomes zero at two
points, viz, $r_{+} = 0.446$ and $r_+=1.739$.
The maximum value for mass is for $r_{+} = 0.872$. Temperature is positive only for the range of points
 $r_{+} \geq 0.872$ and $r_{+} \leq 0.446$. Heat capacity
becomes zero at $r_{+} = 0.446$ and $r_{+} = 0.872$, and suffers an
infinite discontinuity at $r_{+} = 0.59$.

While plotting the temperature and heat capacity verses the horizon radius we could find the so called abnormal behaviours
of the RNdS black hole.

\subsection{Thermodynamic Geometry}

Now we will apply the geometric techniques of Weinhold and Ruppeiner metrics of the system as well as the GTD.
In this case the extensive variables are $N^r =(\alpha, Q)$.
Weinhold metric can be written from the above equation (Eqn[1]) as, 
\begin{equation}
 g^{W}_{ij}=\partial_i \partial_j M(S,Q,\alpha ),
\end{equation}

\begin{eqnarray}
 dS^{2}_{W}= M_{SS} dS^2 +  M_{QQ} dQ^2 + M_{\alpha \alpha} d\alpha^2 + \\ \nonumber
2  M_{SQ} dS dQ + 2  M_{S \alpha} dS d\alpha + 2  M_{ \alpha Q} d\alpha dQ  ,
\end{eqnarray}
and therefore,
   \[
         g^{W}=
            \left[ {\begin{array}{ccc}
             M_{SS} & M_{SQ} & M_{S \alpha} \\
             M_{QS} & M_{QQ} & 0 \\
             M_{\alpha S} & 0 & M_{\alpha \alpha} \\
                \end{array} } \right].
        \]

        The second order partial derivatives can be found using the expression of $M$ given in Eqn[9],
in which $M_{ \alpha Q}$ term will become zero while calculating. So there exists 5 independent elements in the metric. 
 
We could calculate the curvature scalar of the Weinhold metric as, 
\begin{equation}
R^{W}= \frac{\sqrt{S} \alpha^2 \pi^{\frac{3}{2}}(\alpha^2 \pi^2 Q^2+ \alpha^2 \pi S- 9 S^2)}{(\alpha^2 \pi^2 Q^2-\alpha^2 \pi S+ 3 S^2)^2}.
\end{equation}
The numerator is of little physical interest and we are interested in the denominator function. 
The denominator makes the $R^{W}$ singular at \\ $S=\frac{\pi}{6}[\alpha^2\mp\sqrt{\alpha^4 - 12\alpha^2 Q^2}]$.
For each solution of $S$ there exists a pair of $r_+$, which can explain the singularities and zero points in the thermodynamic systems. 
We will avoid the negative values of the solution as it gives imaginary and negative roots. 
For the appropriate choices of the values of the parameter, we can see that it could explain the zero points of heat capacity,
temperature and the peak value of mass at the value $S=2.39$ 
or $r_{+} = 0.872$ and  zeros of temperature and heat capacity at the value $S=0.636$ or $r_{ +} = 0.45$.

Now we will go for the Ruppiner metric which can be conformaly transformed to Weinhold metric. 
Ruppiner metric is given by

   \[
         g^{R}=(\frac{1}{T})
            \left[ {\begin{array}{ccc}
             M_{SS} & M_{SQ} & M_{S \alpha} \\
             M_{QS} & M_{QQ} & 0 \\
             M_{\alpha S} & 0 & M_{\alpha \alpha} \\
                \end{array} } \right],
        \]

which is equal to

\[
         g^{R}=(\frac{4 \alpha^2 \pi^{\frac{3}{2}}S^{\frac{3}{2}}}{\alpha^2 \pi (S - \pi Q^2)- 3 S^2})
            \left[ {\begin{array}{ccc}
             M_{SS} & M_{SQ} & M_{S \alpha} \\
             M_{QS} & M_{QQ} & 0 \\
             M_{\alpha S} & 0 & M_{\alpha \alpha} \\
                \end{array} } \right]
        \]

Now the curvature of the Ruppeiner metric is obtained as, 
\begin{equation}
 R^{R}= \frac{\alpha^2 \pi (2 \pi Q^2-S)}{S(3S^2 - \pi \alpha^2 S+ \pi^2 Q^2 \alpha^2) }.
\end{equation}
 Here also the numerator is of less physical importance. We can see that the denominator throws light on the singularities.
In addition to the singularities obtained from the Weinhold metric, we can see $S=0$ or $r_{+}=0$ is also a singular point in the Ruppiner method. 

We can also use the Legendre invariant transformation of either Weinhold or Ruppiner metric to make further studies.
In GTD it is possible to derive, in principle, an
infinite number of metrics which preserve Legendre invariance,  according to Eqn.[5]. The
simplest way to attain the Legendre invariance for $g^{W}$
is to apply a conformal transformation, with the thermodynamic potential as the conformal factor,
which may unfold other hidden singularities in the present thermodynamic system of black hole.
Here we are taking the conformal transformation which keeps the Legendre invariance as 
\begin{equation}
 g^{W'}= M \otimes g^W.
\end{equation}

Thus we could get the GTD metric as

\[
         g^{W'}=(M)
            \left[ {\begin{array}{ccc}
             M_{SS} & M_{SQ} & M_{S \alpha} \\
             M_{QS} & M_{QQ} & 0 \\
             M_{\alpha S} & 0 & M_{\alpha \alpha} \\
                \end{array} } \right],
        \]

which is equal to

\[
         g^{W'}=(\frac{1}{2} \sqrt{\frac{S}{\pi}} + \frac{Q^2}{2}\sqrt{\frac{\pi}{S}} - \frac{1}{2 \alpha^2} (\frac{S}{\pi})^{\frac{3}{2}})
            \left[ {\begin{array}{ccc}
             M_{SS} & M_{SQ} & M_{S \alpha} \\
             M_{QS} & M_{QQ} & 0 \\
             M_{\alpha S} & 0 & M_{\alpha \alpha} \\
                \end{array} } \right].
        \]

The curvature in the GTD is given by
\begin{equation}
 R^{W'}= \frac{\mathcal{N}}{3 (-S^2+\pi^2 Q^2 \alpha^2 + S \alpha^2 \pi)^3 (3S^2+\pi^2 Q^2 \alpha^2 - S \alpha^2 \pi)^2}.
\end{equation}

Here the numerator($\mathcal{N}$) is a lengthy expression and of little physical importance.  We will get an extra term in the denominator
apart from what we obtained in the Weinhold and Ruppeiner cases. That term makes the $R^{W'}$ singular at
$S = \frac{\pi}{2} [\alpha^{2} \mp \sqrt{\alpha^{4} + 4\alpha^{2} Q^{2}}]$. This can give an extra information about the zero point of mass
at the value $S = 2.39$ or $r_{+} = 0.872$. The singularity in the heat capacity is
not explained here by the curvature.

Finally we will use the most important GTD calculation. In which the choice of thermodynamic potential did not affect the 
answer. In this method we could see the metric using Eqn[5] as follows,

\begin{equation}
         g^{GTD}=(S M_S + Q M_Q + \alpha M_\alpha)
            \left[ {\begin{array}{ccc}
             -M_{SS} & 0 & 0 \\
             0 & M_{QQ} & 0 \\
             0 & 0 & M_{\alpha \alpha} \\
                \end{array} } \right].
        \end{equation}
The corresponding curvature is obtained as
\begin{equation}
  R^{GTD}= \frac{\mathcal{N}}{(3S^2 + \pi \alpha^2 S- 3 \pi^2 Q^2 \alpha^2)^2(S^2 + \pi \alpha^2 S+ 3 \pi^2 Q^2 \alpha^2)^3 }.
\end{equation}

Here also the numerator($\mathcal{N}$) is a lengthy expression and of little physical importance. 
We obtain an extra term in the denominator, and that makes $R^{GTD}$ singular at
$S = \frac{\pi}{6} [-\alpha^{2} \mp \sqrt{\alpha^{4} + 36\alpha^{2} Q^{2}}]$. 
This extra singularity arose here could explain the divergence of heat capacity 
at $S = 1.1$ or $r_{+} = 0.59$.

 \begin{figure}
  \includegraphics[scale=0.8]{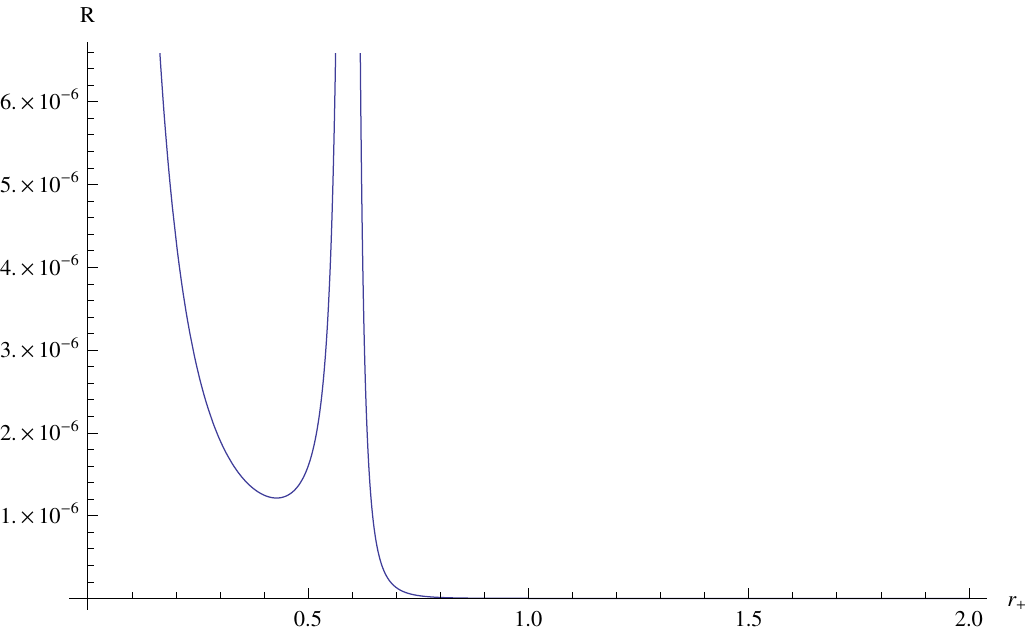}
\caption{\label{fig:orlfo} Variation of most generic curvature scalar of RNdS black hole with respect to horizon radius $r_+$. We set $\alpha=\sqrt{3}$ and $Q=0.4$.}
 \end{figure}

Hence the geometric methods we have elaborated above could give the singularities in the thermodynamic space. 
 The metric which we used above
could be transformed to Legendre invariant metric. The
importance of GTD is that it won't change the results
even if we change the choice of thermodynamic potential
choosen to obtain the metric. Here in our
results we got the singular and zero points repeatedly
in different methods, but the points were unique. And
they are absolutely independent of the choice of thermodynamic potential we used to write the metric.
 
Thus we could plot a most generic curvature scalar
with the horizon radius as in Fig.4. From the figure,
it is easy to find all the singular points and zeros of the
curvature, which uncover the thermodynamic abnormalities in the system. Here the maxima, minima and zeros
of the curvature actually correspond to the thermodynamic behaviour. 
 
Now we are interested in the extremal case of RNdS black hole we discussed above, known as MTZ black hole.

\section{de Sitter black hole coupled with a scalar field}

Now we study the thermodynamics and GTD of dS black hole coupled with a scalar field, which we shall refer to as the MTZ black hole\cite{mtz}. 
MTZ black hole is the solution of the field equations arising from the action

\begin{eqnarray}
 I_L= \int_\mathcal{M} d^4 x \sqrt{-g}[\frac{1}{16 \pi}(R-2 \Lambda ) - \frac{1}{2} g^{ab} \partial_a \phi \partial_b \phi \\ \nonumber
  -\frac{1}{12} R \phi^2 -\zeta \phi^4 -\frac{1}{16 \pi} F^{ab}F_{ab},
\end{eqnarray}

where $\zeta$ is the coupling constant. 
The metric of MTZ black hole is written as
\begin{equation}
 ds^2 = -N(r) dt^2 + N(r)^{-1} dr^2 + r^2 (d\Omega^2),
\end{equation}
 where 
\begin{equation}
 N(r)= \left(1-\frac{M}{r}\right)^2 -\frac{\Lambda}{3} r^2.
\end{equation}
This black hole has inner, event and outer horizons at the values of the radial coordinate $r$  given by

 $r_{-}= \frac{\alpha}{2}\left[-1+ \sqrt{1+\frac{4 M}{\alpha}}\right]$ ;  
 $r_{+}= \frac{\alpha}{2}\left[1- \sqrt{1-\frac{4 M}{\alpha}}\right]$; and 
 $r_{++}= \frac{\alpha}{2}\left[1+ \sqrt{1-\frac{4 M}{\alpha}}\right]$,

where $\alpha= \sqrt{\frac{3}{\Lambda}}$. From the above equation it is clear that, the solution is defined only for $0<M<M_{max}=\frac{\alpha}{4}$. When 
$M=0$ the metric reduces to the de Sitter space and there is only a cosmological horizon. In the other limit , $M=M_{max}=\frac{\alpha}{4}$, 
the event and the cosmological horizons coincide, leaving the charge at the Nariai limit. 
 Now we can do the usual thermodynamics of the event horizon at $r_+$.

\subsection{Thermodynamics}
 The usual thermodynamics of MTZ black hole is explained below.
 Using area law, we can express entropy as a function of $M$ and $\alpha$ as, 
\begin{equation}
 S(M,\alpha)=\frac{\pi \alpha^2}{4} \left[1-\sqrt{1-\frac{4 M}{\alpha}}\right]^2.
\end{equation}
We can now deduce the mass as,
\begin{equation}
 M=  \frac{S}{\alpha \pi}-\sqrt{\frac{S}{\pi}}
\end{equation}

 \begin{figure}[h]
\includegraphics[scale=0.8]{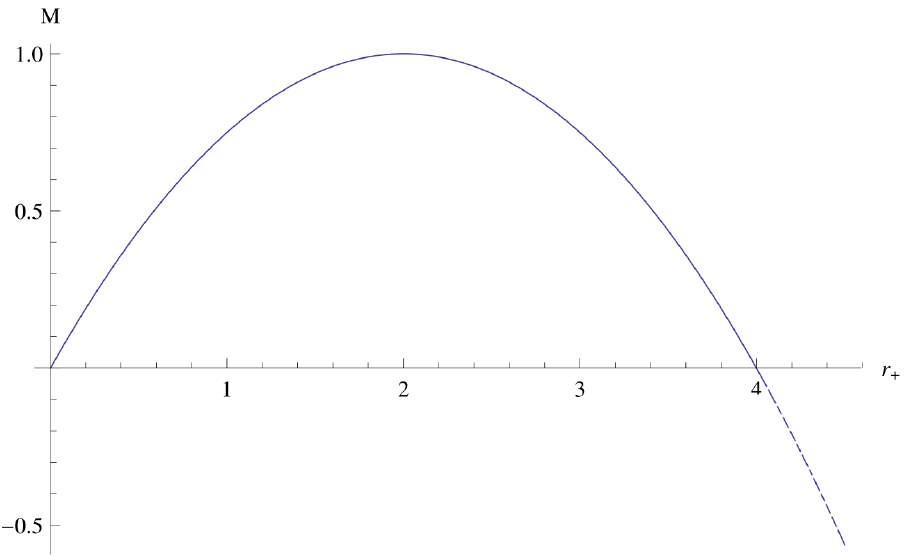}
\caption{\label{fig:orlfo} Variation of mass with respect to horizon radius $r_+$. We set $\alpha=4$. The dashed portion is the negative - unphysical regime of mass.}
\end{figure}

\begin{figure}[h]
\includegraphics[scale=0.8]{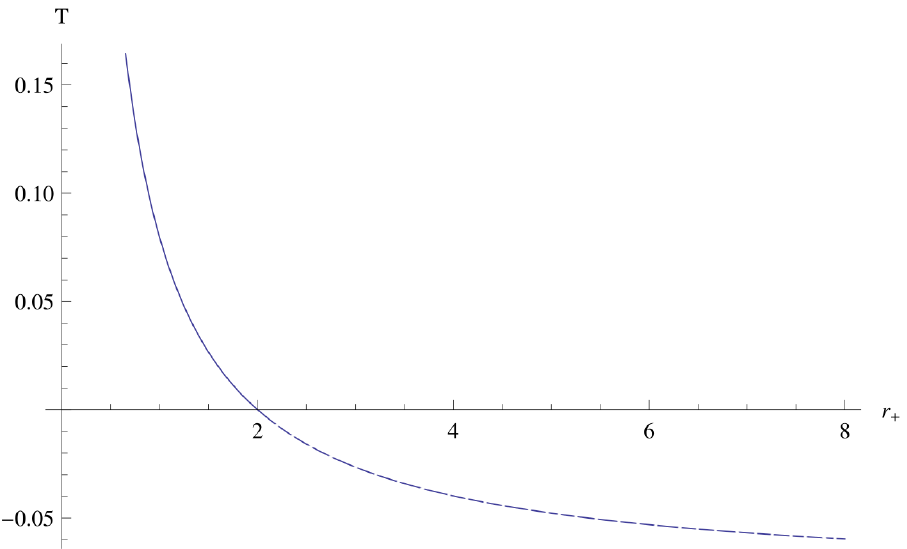}
\caption{\label{fig:orlfo} Variation of temperature with respect to horizon radius $r_+$. We set $\alpha=4$. The dashed portion is the negative - unphysical regime of temerature.}
\end{figure}

Now we can straightforwardly write the temperature 
as

\begin{equation}
 T=  \frac{1}{\alpha \pi}-{\frac{1}{2\sqrt{\pi S}}},
\end{equation}

 while the  heat capacity is 
\begin{equation}
C=-2S+ \frac{4 S^{\frac{3}{2}}}{\alpha \sqrt{\pi}}.
\end{equation}

\begin{figure}[h]
\includegraphics[scale=0.8]{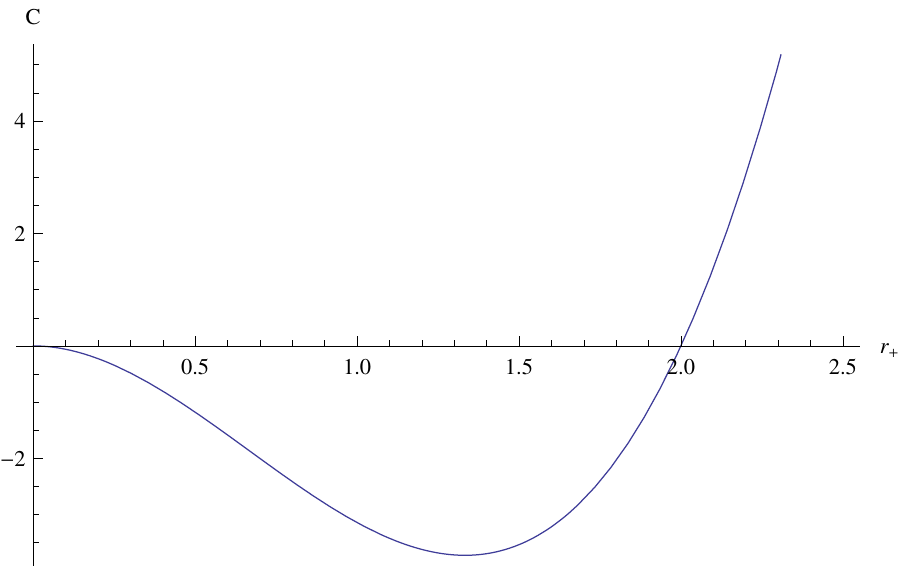}
\caption{\label{fig:orlfo} Variation of heat capacity with respect to horizon radius $r_+$. We set $\alpha=4$.}
\end{figure}

Here we have plotted all the three thermodynamic parameters as a function of horizon radius($r_{+}$).  In Fig.5, we find
that mass of the black hole is positive only for a particular range of $r_+$. In the
case of temperature also, in Fig.6, it goes to the negative range
after a particular value of $r_+$. For heat capacity, here there
is no infinite discontinuity, but it also falls to the
negative(unstable) phase after a particular value of $r_+$, as in Fig.7.

Thus we see that the thermodynamic parameters of the MTZ black hole also shows abnormal behaviours.
The particular point about the present system is the validity of the choice of $r_+$, which has been used to find the horizon thermodynamics here.
The value of $r_+$ lies in the range, $M<r_+<\frac{\alpha}{2}$.

Mass of the black hole becomes zero at $r_+=0$ and at $r_+=4$. $r_+=2$  corresponds to the maxima of the curve in Fig.5. 
Temperature of the black hole becomes negative after reaching zero at $r_+=2$. Heat capacity also reaches zero at $r_+=2$ 
and then falls to the unstable phase.

\subsection{Thermodynamic geometry}

Now we construct the thermodynamic geometry of the MTZ black hole using Weinhold metric.
In this case the extensive variable is $N^r =[\alpha]$ so that the general Weinhold metric becomes,

\begin{equation}
         g^{W}=
            \left[ {\begin{array}{cc}
             M_{SS} & M_{S \alpha} \\
             M_{\alpha S} & M_{\alpha \alpha}\\
                \end{array} } \right].
        \end{equation}
We could easily find the metric elements from the expression of mass given by Eqn[23] 
and the corresponding scalar curvature becomes
\begin{equation}
 R^W=-\frac{\alpha^2 \pi^{\frac{3}{2}}}{\left[2\sqrt{S}(\alpha \sqrt{\pi}- 2 \sqrt{S})^2 \right]}.
\end{equation}
This curvature could explain two singularities in the thermodynamics of MTZ black hole. One at which 
$S=0$ and the other at which $S=\frac{\alpha^2 \pi}{4}$. The latter brings the information of zeros in both heat capacity and
temperature. 

Now we construct the Ruppeiner metric, 
\begin{equation}
g^{R}= (-\frac{1}{T})
            \left[ {\begin{array}{cc}
             M_{SS} & M_{S \alpha} \\
             M_{\alpha S} & M_{\alpha \alpha}\\
                \end{array} } \right],
        \end{equation}
while calculating the curvature scalar, we can see that it is zero. So we can't explain any of the singular behavior of the
thermodynamic system under consideration using Ruppeiner method. 
Now we apply the GTD approach and we can find the GTD metric in a couple of ways by Legendre invariant conformal transformations,
\begin{equation}
g^{GTD}= (SM_S + \alpha M_\alpha)
            \left[ {\begin{array}{cc}
            - M_{SS} & 0 \\
             0& M_{\alpha \alpha}\\
                \end{array} } \right]
        \end{equation}
 and the corresponding  curvature is
\begin{equation}
  R^{GTD}=-\frac{3\pi}{S},
\end{equation}
which obviously gives only one singularity. 

Another Legendre transforms of this as $g\rightarrow \Lambda^{-1} g$
gives 

\begin{equation}
g^{GTD*}= (SM_S + \alpha M_\alpha)^{-1}
            \left[ {\begin{array}{cc}
            - M_{SS} & 0 \\
             0& M_{\alpha \alpha}\\
                \end{array} } \right]
        \end{equation}

Here the curvature gives 
\begin{equation}
 R^{GTD*}= -\frac{1}{4}.
\end{equation}

It's merely a scalar and independent of any of the parameters. So in this case, the Weinhold metric gives more good results regarding the thermodynamic
abnormalities of MTZ black hole. Thus we consider a generalized Weinhold metric as in
Eqn[16] as,
\begin{equation}
         g^{W}=(M)
            \left[ {\begin{array}{cc}
             M_{SS} & M_{S \alpha} \\
             M_{\alpha S} & M_{\alpha \alpha}\\
                \end{array} } \right].
        \end{equation}
Following the previous steps, we could see that the curvature scalar gets the form,

\begin{equation}
 R^{W*}=\frac{\alpha^3 \pi^{\frac{5}{2}}(3 \alpha^2 \pi - 9 \alpha \sqrt{\pi S}+ 8S)}{4S(\alpha \sqrt{\pi}- 2 \sqrt{S})^2 (\sqrt{S}-\alpha \sqrt{\pi})^3}.
\end{equation}

\begin{figure}
  \includegraphics[scale=0.8]{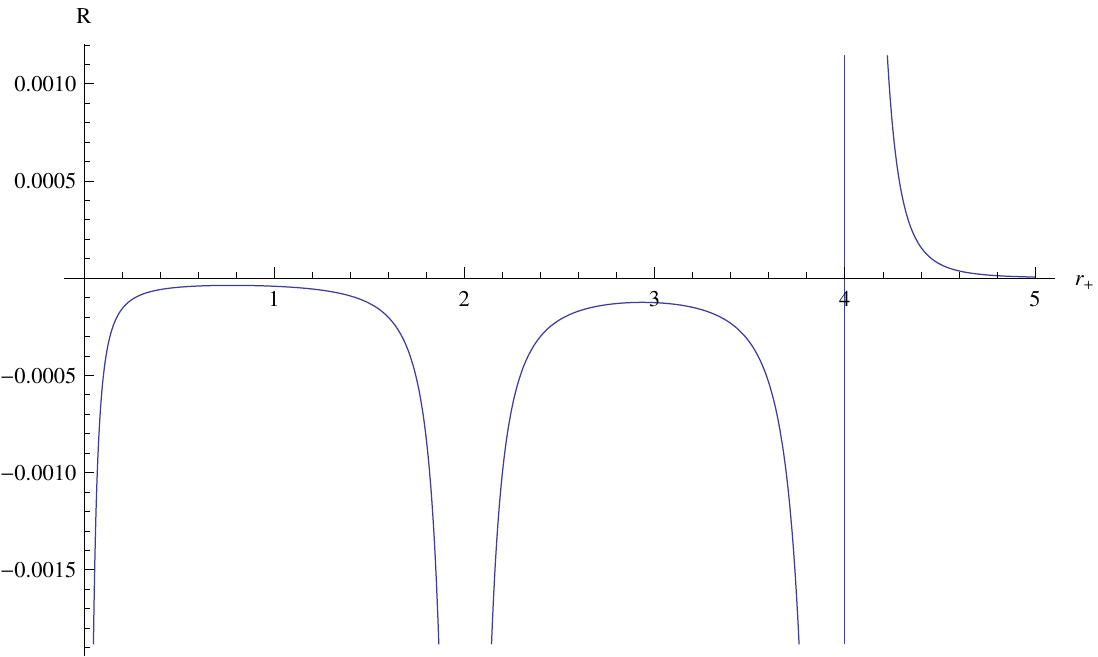}
\caption{\label{fig:orlfo} Variation of most generic curvature scalar of MTZ black hole with respect to horizon radius $r_+$. We set $\alpha=4$.}
 \end{figure}
 
Here the generalized Weinhold metric reveals the remaining zeros and singularities of the present thermodynamic
system. One singularity is obviously at $S = 0$ or $r_{ +} = 0$; where the mass and heat capacity have zero values. 
Second singularity is at $S = \frac{\alpha^{2} \pi}{4}$ or $r_{+} =2$; where the temperature and heat capacity have  
zero value, and mass has the maximum value.The final singularity is at $S = \pi \alpha^{2}$ which in turn gives $r_{+} = 4$ and this 
corresponds to the second point in Fig. 5, where mass becomes zero. Thus, this
metric provides us almost all of the thermodynamic singular and zero points. 
In general, the geometric methods uniquely determine the points of the thermodynamic parameters of MTZ black hole at which the
thermodynamic quantities become singular. The generic scalar curvature
can be plotted with horizon radius as in Fig.8, and
all the thermodynamic behaviours are revealed in this calculation.

\section{Conclusion}

In this work we have analyzed the thermodynamics and thermodynamic geometry of RNdS black hole and its extremal case. 
First we have found the thermodynamics of RNdS black hole. The thermodynamics showed abnormalities in
temperature, mass and heat capacity.
The heat capacity exhibits a second order phase transition as well. So our aim is to express these abnormalities
with the geometric tools available in thermodynamics. We have used three different thermodynamic geometric
techniques to analyze these abnormalities.

 We have found that in the case of RNdS black hole,  the geometric methods of Weinhold, Ruppeiner and Quevedo(GTD), combined together give
 the singularities which could explain the abnormal behaviours of the system  whereas in the the case of MTZ black hole, the generalized Weinhold method only gives
 the correct results.

\section{Acknowledgments}

We thank the reviewer for the useful comments.
We are grateful to Hernando Quevedo for the fruitful discussions. 
TR wishes to thank UGC, New Delhi for financial support
under RFSMS scheme. JS wishes to thank UGC for the project under which  he is working.  VCK is thankful to UGC, New Delhi for financial
support through a Major Research Project and wishes to acknowledge
Associateship of IUCAA, Pune, India.

\end{document}